\documentclass[english,prb, final, twocolumn, showpacs]{revtex4}
\usepackage{lmodern}
\usepackage[T1]{fontenc}
\usepackage[latin9]{inputenc}
\usepackage{verbatim}
\usepackage{amsmath}
\usepackage{graphicx}
\usepackage{amssymb}

\makeatletter
\@ifundefined{textcolor}{}
{%
 \definecolor{BLACK}{gray}{0}
 \definecolor{WHITE}{gray}{1}
 \definecolor{RED}{rgb}{1,0,0}
 \definecolor{GREEN}{rgb}{0,1,0}
 \definecolor{BLUE}{rgb}{0,0,1}
 \definecolor{CYAN}{cmyk}{1,0,0,0}
 \definecolor{MAGENTA}{cmyk}{0,1,0,0}
 \definecolor{YELLOW}{cmyk}{0,0,1,0}
 }

\usepackage{hyperref}

\makeatother

\usepackage{babel}

\begin{document}

\title{Dissipation-driven superconductor-insulator transition in linear
arrays of Josephson junctions capacitively coupled to metallic films}
\begin{abstract}
We study the low-temperature properties of linear Josephson-junction
arrays capacitively coupled to a proximate two-dimensional diffusive
metal. Using bosonization techniques, we derive an effective model
for the array and obtain its critical properties and phases at $T=0$
using a renormalization group analysis and a variational approach.
While static screening effects given by the presence of the metal
can be absorbed in a renormalization of the parameters of the array,
backscattering originated in the dynamically screened Coulomb interaction
produces a non-trivial stabilization of the insulating groundstate
and can drive a superconductor-insulator transition. We study the
consequences for the transport properties in the low-temperature regime.
In particular, we calculate the resisitivity as a function of the
temperature and the parameters of the array, and obtain clear signatures
of a superconductor-insulator transition that could be observed in
experiments. 
\end{abstract}

\author{Alejandro M. Lobos}

\affiliation{DPMC-MaNEP, University of Geneva, 24 Quai Ernest-Ansermet CH-1211
Geneva, Switzerland.}

\author{Thierry Giamarchi}

\affiliation{DPMC-MaNEP, University of Geneva, 24 Quai Ernest-Ansermet CH-1211
Geneva, Switzerland.}

\date{\today}


\maketitle

\section{Introduction}

Low-dimensional superconductors are systems displaying a surprisingly
complex and rich physics, allowing the study of paradigmatical phenomena
in condensed matter physics, like quantum phase transitions and quantum
critical behavior, electronic localization, Coulomb blockade, etc.
\citep{schoen90_review_ultrasmall_tunnel_junctions,fazio01_review_superconducting_networks}
In particular, an intriguing superconductor-insulator phase transition
(SIT) was observed experimentally in superconducting films \citep{Haviland89_Onset_of_superconductivity_in_the_two-dimensional_limit_PhysRevLett.62.2180,Mason99_Dissipation_effects_on_the_SIT_in_2D,Mason02_SIT_in_a_capacitively_coupled_dissipative_environment},
wires \citep{bezryadin00,arutyunov08_superconductivity_1d_review,Bezryadin08_QPS_review},
and in ultrasmall-capacitance Josephson junction arrays (JJAs) in
two \citep{vanderZant96_QPT_in_2D_Experiments_in_JJas,rimberg97_dissipation_driven_sit_2D_josephson_array,Takahide00_SIT_in_2DJJAs}
and one dimensions \citep{Chow98_Length_scale_dependence_of_SIT_in_1D_array_J,Kuo01_Magnetic_induced_transition_in1DJJA,Miyazaki02_QPT_in_1D_arrays_of_JJs,Takahide06_SIT_2D_1D_crossover},
giving rise to an intense theoretical activity.\citep{Fisher1989,Fisher1990}
In this transition, as one of the parameters is varied (e.g., the
normal-state resistance of the film, the thickness of the wire, the
Josephson coupling $E_{J}$ in the array, etc.) the groundstate of
the system changes from superconducting to insulator. 

In one-dimensional (1D) superconductors, one particular kind of excitation,
the so-called quantum-phase slip (QPS) processes, have been recently
the focus of an intense research.\citep{Bezryadin08_QPS_review} The
interest is based both on the putative role of QPS in the SIT in 1D\citep{zaikin97},
as well as for their potential uses in novel qubit architectures\citep{Mooij05_Phase-slip_qubit},
a fact that has stimulated recent interesting experimental research
in 1DJJAs.\citep{Pop10_QPS_in_JJA,Manucharyan10_Coherent_QPS_in_JJA}
A phase-slip is a discrete process occuring in a 1D superconductor,
in which the amplitude of the order parameter vanishes temporarily
at a particular point, allowing the phase of the order parameter to
change abruptly in units of $2\pi$. In particular, a\textit{ }QPS
is a phase-slip excitation originated in macroscopic quantum tunneling
of the phase of the order parameter. \citep{giordano94}

On the other hand, since the seminal works by Caldeira and Leggett\citep{caldeira&leggett81},
it has been known that dissipation in macroscopic quantum systems
plays a central role. For instance, in a two-dimensional JJA capacitively
coupled to a proximate two-dimensional electron gas (2DEG), Rimberg
\textit{et al.} observed a tunable-SIT upon variation of the backgate
voltage $V_{g}$ applied to the 2DEG.\citep{rimberg97_dissipation_driven_sit_2D_josephson_array}
In that work, it was shown that $V_{g}$ has the effect of tuning
the sheet resistance $R_{\square}$ in the 2DEG through the modulation
of its electronic density, a fact that in turn modifies the electromagnetic
environment of the JJA. It was argued later by Wagenblast \textit{et
al.} \citep{Wagenblast97_SIT_in_a_Tunable_Dissipative_Environment}
that due to the incomplete screening of the Coulomb interaction provided
by the 2DEG, the 2D plasma mode in the array is overdamped and the
charging energy in the junction $E_{C}$ is renormalized to higher
values, producing a SIT whenever the ratio $E_{J}/\tilde{E}_{C}\sim1$,
with $\tilde{E}_{C}$ the renormalized $E_{C}$. While this scenario
is reasonable in a 2D geometry, in a 1DJJA capacitively coupled to
a 2DEG, the screening provided by the metal is typically very efficient,
and a significant damping of the 1D propagating plasma mode is not
expected.\citep{Lobos10_Dissipative_phase_fluctuations} This leads
to the naive conclusion that in the 1DJJA/2DEG geometry, a dissipation-driven
SIT should not occur. %
\begin{figure}[h]
\includegraphics[bb=75bp 50bp 700bp 270bp,clip,scale=0.4]{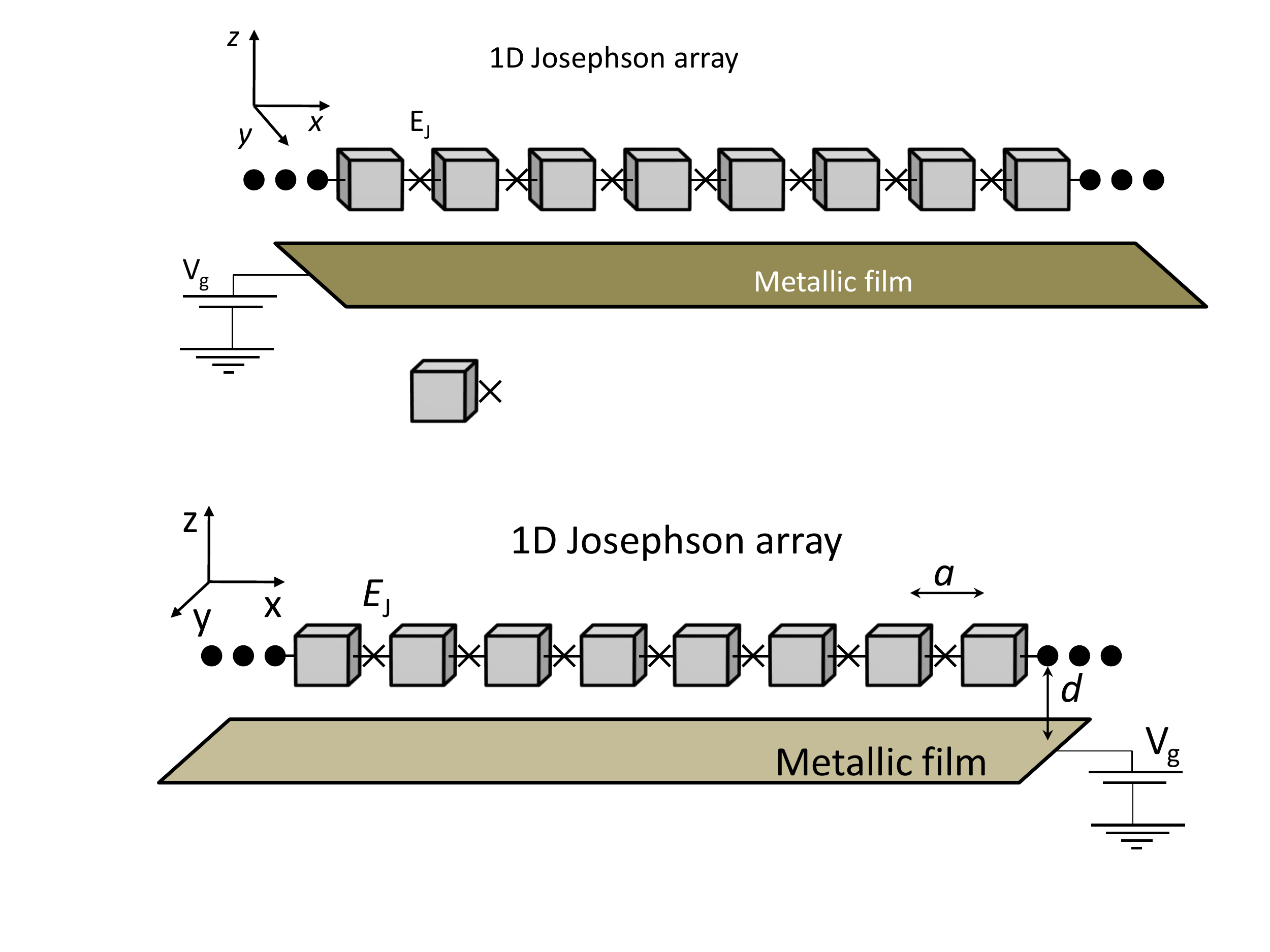}\caption{\label{fig:system}Scheme of the system under study. The 1DJJA is
capacitively coupled to the metallic film, which controls the electromagnetic
environment. A gate voltage allows to modify the sheet-resistance
$R_{\square}$ of the film, modifying the dissipation in the 1DJJA.}

\end{figure}
However, in a closely related Luttinger-liquid system placed in proximity
to a metallic plane, a dissipation-driven quantum phase transition
was predicted to occur.\citep{cazalilla06_dissipative_transition}
This transition is driven by backscattering events originated in the
Luttinger liquid under the effect of the dynamically screened Coulomb
interaction. It is therefore interesting to study to what extent the
same dissipative processes will affect the dynamics of QPS in 1D superconductors
in proximity of a diffusive metallic plane. Indeed, the question of
dissipation in 1D superconductors is an active area of research, and
theoretical predictions point towards the important role of intrinsic
and/or extrinsic dissipation mechanisms in determining their $T=0$
phase diagram.\citep{goswami06_josephson_array,refael07_SN_transition_in_grains&nanowires}

In this article we explore the possibility of a dissipation-driven
SIT in a 1DJJA capacitively coupled to a proximate diffusive 2DEG.
We concentrate in particular on the low-temperature phase diagram
and on the transport properties of the array. Using a bosonization
approach,\citep{giamarchi_book_1d} we derive the dissipative effective
action from a microscopic Hamiltonian, and we elucidate the role of
dissipation in the SIT. One important conclusion in our work is that
for weak dissipation, the transition occurs always between a superconducting
and an insulating phase, in contrast to other works predicting quadrupolar
and normal phases.\citep{goswami06_josephson_array,refael07_SN_transition_in_grains&nanowires}
We believe this is a consequence of a different kind of dissipation
in the model. We find that, except in the experimentally challenging
situation in which the Cooper-pair density in the superconducting
islands is not commensurate, the SIT is always of the Berezinskii-Kosterlitz-Thouless
(BKT) type,\citep{Berezinskii1971,Kosterlitz1973} and is originated
in the unbinding of QPS/anti-QPS pairs. Dissipation stabilizes the
insulating groundstate through the introduction of friction in the
dynamics of the 1D superfluid density, a fact that could be observed
experimentally in the dc-resistivity of the 1DJJA. Specifically, we
predict a resistivity of the form $\varrho\left(T\right)\sim A_{1}T^{\nu_{1}}+A_{2}T^{\nu_{2}}$
in the superconducting phase, and $\varrho\left(T\right)\sim\eta Te^{\Delta/T}/\Delta$,
in the insulating phase, with $\Delta$ the insulating gap and $\eta$
the dissipation parameter.

The paper is organized as follows. In Section \ref{sec:model} we
derive the effective model for a 1DJJA coupled to a 2DEG, in Section
 \ref{sec:phase-diagram} we derive the $T=0$ phase-diagram as a
function of the parameters of the model, Section \ref{sec:transport}
is devoted to the study of the experimental consequences of our results,
and finally in Section \ref{sec:summary} we present a summary and
our conclusions.

\section{\label{sec:model}Model }

We start the analysis by considering an ideally isolated JJA, with
length $L\rightarrow\infty$. To simplify the analysis, we neglect
in the following the fermionic degrees of freedom forming the Cooper-pairs
at a microscopic level. This {}``boson-only'' approximation is belived
to describe correctly the critical properties of a JJA at temperatures
$T\ll T_{c}$, with $T_{c}$ the superconducting critical temperature
in the bulk of the superconducting island.\citep{Fisher1989} The
usual description of the isolated, infinite 1DJJA is given in terms
of the quantum phase model \citep{fazio01_review_superconducting_networks}\begin{align}
H_{\text{JJA}} & =\frac{1}{2}\sum_{i,j}\left(n_{i}-\bar{n}\right)v_{ij}\left(n_{j}-\bar{n}\right)+\sum_{\left\langle ij\right\rangle }E_{J}\left(1-\cos\theta_{i}-\theta_{j}\right).\label{eq:H_JJA}\end{align}
The dynamical variables of this model are the number of Cooper pairs
$n_{i}$ and the phase of the superconducting order parameter $\theta_{i}$
at every site $i$ in the array. These variables obey the usual phase-number
commutation relation in the BCS groundstate, i.e., $\left[\theta_{i},n_{j}\right]=i\delta_{ij}.$\citep{Tinkham}
The first term in Eq. (\ref{eq:H_JJA}) represents the charging energy,
with $v_{ij}$ the unscreened Coulomb interaction {[}cf. Eq. (\ref{eq:Coulomb_potential}){]}
between the excess charges at sites $i$ and $j$, and $\bar{n}$
corresponds to an average charge imposed, e.g., by external gate voltages.
The second term is the Josephson energy contribution, parametrized
by $E_{J}$. In the following we use the convention $\hbar=k_{B}=1$.

The critical properties of model Eq. (\ref{eq:H_JJA}) are more conveniently
studied using a field-theoretical approach, valid for fluctuations
of wavelengths much larger that the lattice parameter of the array
$a$.\citep{giamarchi_book_1d} We therefore introduce the coarse-grained
superfluid density $\delta\rho\left(x\right)$, defined as $\delta\rho\left(x_{i}\right)=\left(n_{i}-\bar{n}\right)/a$,
and we expand the Josephson term as $E_{J}\cos\left(\theta_{i}-\theta_{j}\right)\simeq E_{J}a^{2}\left[\nabla\theta\left(x_{i}\right)\right]^{2}$.
At low temperatures, the continuum limit of Hamiltonian Eq. (\ref{eq:H_JJA})
reads\begin{align}
H_{\text{JJA}} & =\frac{1}{2}\int dxdx^{\prime}\;\delta\rho\left(x\right)v\left(x-x^{\prime},0\right)\delta\rho\left(x^{\prime}\right)\nonumber \\
 & +\frac{1}{2}E_{J}a\int dx\;\left(\nabla\theta\left(x\right)\right)^{2}.\label{eq:H_JJA_continuum}\end{align}
Here the 1D superfluid interacts via the \textit{bare} Coulomb potential,
which we define for convenience as \begin{align}
v\left(\mathbf{r},z\right) & =\frac{e^{2}}{\epsilon_{\text{r}}}\frac{1}{\sqrt{r^{2}+z^{2}+a^{2}}},\label{eq:Coulomb_potential}\end{align}
where $r=\left|\mathbf{r}\right|$ and $z$ are, respectively, the
distance in the $xy$-plane and along the azimuthal direction between
two point-charges (cf. Fig. \ref{fig:system}). Here the lattice parameter
$a$ acts as the short-distance regularization of the interaction
and $\epsilon_{\text{r}}$ is the permitivity of the insulating medium
surrounding the islands. Note that in Eq. (\ref{eq:H_JJA_continuum})
we do not assume an \textit{a priori} short-ranged, screened interaction
as is usually done when dealing with JJAs.\citep{Wagenblast97_SIT_in_a_Tunable_Dissipative_Environment}
This will result as a natural consequence of the interaction with
the 2DEG (see below). One problem of this field-theoretical approach
is that Mott-instabilities (crucial when the superfluid density is
commensurate with the lattice) are lost in Eq. (\ref{eq:H_JJA_continuum})
after taking the continuum limit. One way to cure this problem is
to introduce a phenomenological term $H_{1}=-\int dx\; V_{l}\left(x\right)\rho\left(x\right)$,
where the effective superfluid density $\rho\left(x\right)$ {[}cf.
Eq. (\ref{eq:density_bosonized}){]} couples to the phenomenological
potential $V_{l}\left(x\right)$, having the same periodicity of the
lattice.\citep{giamarchi_book_1d}

The electrons in the 2DEG are described by the Hamiltonian\begin{align}
H_{\textrm{2D}} & =\int d^{2}\mathbf{r}\;\sum_{\sigma}\left[-\frac{1}{2m}\eta_{\sigma}^{\dagger}\nabla^{2}\eta_{\sigma}+V_{\textrm{imp}}\eta_{\sigma}^{\dagger}\eta_{\sigma}\right]+\nonumber \\
 & +\frac{1}{2}\int d^{2}\mathbf{r}d^{2}\mathbf{r}^{\prime}\;\delta\rho_{\textrm{2D}}\left(\mathbf{r}\right)v\left(\mathbf{r}-\mathbf{r}^{\prime},0\right)\delta\rho_{\textrm{2D}}\left(\mathbf{r}^{\prime}\right),\label{eq:h_n}\end{align}
where the fermionic field-operator $\eta_{\sigma}^{\dagger}\equiv\eta_{\sigma}^{\dagger}\left(\mathbf{r}\right)$
creates an electron in the 2DEG with spin projection $\sigma$ at
spatial position $\mathbf{r}\equiv\left(x,y\right)$, and $V_{\textrm{imp}}\equiv V_{\textrm{imp}}\left(\mathbf{r}\right)$
represents a weak static impurity potential which provides a finite
resitivity and dissipation in the metal. In terms of $\eta_{\sigma}^{\dagger}\left(\mathbf{r}\right),\eta_{\sigma}\left(\mathbf{r}\right)$,
the density-operator $\rho_{\textrm{2D}}\left(\mathbf{r}\right)$
in the 2DEG writes $\rho_{\textrm{2D}}\left(\mathbf{r}\right)\equiv\sum_{\sigma}\eta_{\sigma}^{\dagger}\left(\mathbf{r}\right)\eta_{\sigma}\left(\mathbf{r}\right)$,
and $\delta\rho_{\textrm{2D}}\left(\mathbf{r}\right)\equiv\rho_{\textrm{2D}}\left(\mathbf{r}\right)-\rho_{0,\textrm{2D}}$,
with $\rho_{0,\textrm{2D}}$ the average density in the metal. 

Finally, the interaction between the 1DJJA and the 2DEG placed at
a distance $d$ (cf. Fig. \ref{fig:system}) is described by the Hamiltonian\begin{align}
H_{\textrm{int}} & =\int d^{2}\mathbf{r}dx^{\prime}\;\delta\rho\left(x^{\prime}\right)v\left(x^{\prime}-\mathbf{r},d\right)\delta\rho_{\textrm{2D}}\left(\mathbf{r}\right).\label{eq:h_int}\end{align}

Our goal in this Section is to derive an effective model for the 1DJJA
capacitively coupled to the 2DEG. To that end we introduce the partition
function of the system\citep{negele_book}

\begin{align*}
Z & =\int\mathcal{D}\left[\rho,\theta\right]\mathcal{D}\left[\bar{\eta},\eta\right]\; e^{-S},\end{align*}
where $S$ is the Euclidean action of the problem\begin{align}
S & =S_{\textrm{JJA}}+S_{\textrm{2D}}+S_{\textrm{int}},\label{eq:S_eff_total}\end{align}
where \begin{align*}
S_{\text{JJA}} & =\int_{0}^{\beta}d\tau\int dx\; i\partial_{\tau}\theta\left(x,\tau\right)\rho\left(x,\tau\right)+\int_{0}^{\beta}d\tau\; H_{\text{JJA}}\left(\tau\right),\\
S_{\text{2D}} & =\int_{0}^{\beta}d\tau\left[\int d^{2}\mathbf{r}\;\bar{\eta}\left(\mathbf{r},\tau\right)\left(\partial_{\tau}-\mu_{\text{2D}}\right)\eta\left(\mathbf{r},\tau\right)+H_{\text{2D}}\left(\tau\right)\right],\\
S_{\text{int}} & =\int_{0}^{\beta}d\tau\; H_{\text{int}}\left(\tau\right).\end{align*}
Here $\mu_{\textrm{2D}}=k_{\text{F}}^{2}/2m-eV_{g}$ is the effective
chemical potential in the metal, with $k_{\text{F}}=\left|\mathbf{k}_{\text{F}}\right|$
the Fermi wavevector, and $V_{g}$ the gate voltage applied to the
2DEG, which allows to change the value of $\rho_{0,\textrm{2D}}$,
and therefore, the sheet-resistance $R_{\square}$. 

The first step in the derivation of an effective model for the array
is to integrate out the fermionic degrees of freedom $\bar{\eta}\left(\mathbf{r},\tau\right),\eta\left(\mathbf{r},\tau\right)$
in the 2DEG. Assuming that the term $S_{\textrm{int}}$ can be treated
perturbatively (we check the consistency of this assumption later),
the integration of the fermionic degrees of freedom in the metal yields
\begin{align}
S_{\textrm{eff}} & \simeq S_{\textrm{JJA}}-\frac{1}{2}\int d\tau d\tau^{\prime}\int dxdx^{\prime}\;\delta\rho\left(x,\tau\right)\nonumber \\
 & \times v_{\textrm{scr}}\left(x-x^{\prime},\tau-\tau^{\prime}\right)\delta\rho\left(x^{\prime},\tau^{\prime}\right).\label{eq:S_2nd_order}\end{align}
We do not provide the details of this derivation here, and we refer
the interested reader to Refs. \onlinecite{Lobos10_Dissipative_phase_fluctuations}
and \onlinecite{cazalilla06_dissipative_transition}. In Eq. (\ref{eq:S_2nd_order})
we have introduced the 1D effective screening potential $v_{\textrm{scr}}\left(x-x^{\prime},\tau-\tau^{\prime}\right)$,
which encodes all the screening effects provided by the 2DEG. This
quantity writes more conveniently in Fourier representation as\citep{Lobos10_Dissipative_phase_fluctuations}
\begin{align}
v_{\textrm{scr}}\left(k,\omega_{m}\right) & \equiv\frac{1}{L_{\perp}}\sum_{k_{\perp}}\frac{\left[v_{\textrm{2D}}\left(\mathbf{k},d\right)\right]^{2}\chi_{0,\textrm{2D}}\left(\mathbf{k},\omega_{m}\right)}{1+v_{\textrm{2D}}\left(\mathbf{k},0\right)\chi_{0,\textrm{2D}}\left(\mathbf{k},\omega_{m}\right)},\label{eq:v_scr}\end{align}
where $\omega_{m}=2\pi m/\beta$ are the bosonic Matsubara frequencies,\citep{mahan2000}
and $\mathbf{k}=\left(k,k_{\perp}\right)$ is the wavevector in 2D,
where we have made explicit the component $k_{\perp}$ in the 2DEG,
perpendicular to the 1DJJA. The quantity $v_{\textrm{2D}}\left(\mathbf{k},d\right)=\left(2\pi e^{2}/\epsilon_{\text{r}}\right)\exp\left(-\left|\mathbf{k}\right|\sqrt{d^{2}+a^{2}}\right)/\left|\mathbf{k}\right|$
is the 2D Fourier transform of the Coulomb potential Eq. (\ref{eq:Coulomb_potential}).
We assume that the length of the array is $L<\xi_{\text{loc}}$, with
$\xi_{\text{loc}}$ the Anderson localization length in the 2DEG,
a condition well fulfilled in practice. In that case, the density-density
response function in the 2DEG, averaged over disorder configurations,
writes $\chi_{0,\textrm{2D}}\left(\mathbf{k},\omega_{m}\right)=2\mathcal{N}_{2D}^{0}D\mathbf{k}^{2}/\left(D\mathbf{k}^{2}+\left|\omega_{m}\right|\right)$,
where $D$ and $\mathcal{N}_{2D}^{0}$ are, respectively, the diffusion
constant and the density of states (at the Fermi energy) per spin
projection.\citep{akkermans} 

We can now define the total effective retarded interaction

\begin{align}
v_{\textrm{eff}}\left(k,\omega_{m}\right) & =v_{\textrm{1D}}\left(k,0\right)-v_{\textrm{scr}}\left(k,\omega_{m}\right),\label{eq:v_eff}\end{align}
where $v_{\textrm{1D}}\left(k,0\right)=2e^{2}K_{0}\left(\left|k\right|a\right)/\epsilon_{\text{r}}$
is the Fourier transform of Eq. (\ref{eq:Coulomb_potential}) in 1D,
and $K_{0}\left(x\right)$ is the zeroth-order modified Bessel function.\citep{abramowitz_math_functions}
Physically, the effective potential $v_{\textrm{eff}}\left(k,\omega_{m}\right)$
describes the interaction among charges in the array, both via the
instantaneous interaction $v_{\textrm{1D}}\left(k,0\right)$ arising
from the direct intrawire Coulomb interaction, as well as indirectly
via the coupling to the diffusive modes in the 2DEG, which corresponds
to the retarded interaction $v_{\textrm{scr}}\left(k,\omega_{m}\right)$
Eq. (\ref{eq:v_scr}).

We now introduce a more convenient representation of the superfluid
density in the 1DJJA. To motivate our approach, we first note that
in the absence of Josephson coupling {[}i.e., $E_{J}=0$ in Eq. (\ref{eq:H_JJA}){]},
the Cooper-pair occupation number $n_{i}$ is a good quantum number
in each island, fixed by $\bar{n}$ via the application of an external
gate-voltage. Increasing $E_{J}$ will evidently introduce fluctuations
in $n_{i}$ due to the transfer of Cooper-pairs between neighboring
islands, and $n_{i}$ is no longer a good quantum number. However,
we expect that in the experimentally interesting regime $E_{J}/E_{0}\sim1$,
where $E_{0}$ is the characteristic charging energy in the island,
fluctuations $\Delta n_{i}\equiv n_{i}-\bar{n}$ will be of order
$\Delta n_{i}\simeq\pm1$, and that all other charging states such
that $\left|\Delta n_{i}\right|\gg1$ will be energetically forbiden.
We therefore truncate those states from our description and focus
on charge-fluctuations of $\Delta n_{i}=\pm1$. In terms of a continuous
field $\phi\left(x\right)$, which is slowly varying on the scale
of $a$, the superfluid density in this effective model can be more
conveniently written as\citep{Haldane1981}

\begin{align}
\rho\left(x\right) & =\left[\rho_{0}-\frac{1}{\pi}\nabla\phi\left(x\right)\right]\sum_{p}e^{i2p\left(\pi\rho_{0}x-\phi\left(x\right)\right)},\label{eq:density_bosonized}\end{align}
where the parameter $\rho_{0}$ is defined as $\rho_{0}\equiv1/a$
in the commensurate case. Note that $\rho_{0}$ is an effective parameter
of our model, and cannot be interpreted as the total physical density,
in contrast to truly 1D systems.\citep{giamarchi_book_1d} Only the
\textit{fluctuations} $\delta\rho\left(x\right)\equiv\rho\left(x\right)-\rho_{0}$
have a physical meaning in our model. 

In order to obey the phase-number commutation relations in the BCS-groundstate,\citep{Tinkham}
note that the field $\phi\left(x\right)$ must verify the new commutation
relation\begin{align}
\left[\theta\left(x\right),\nabla\phi\left(x^{\prime}\right)\right] & =i\pi\delta\left(x-x^{\prime}\right).\label{eq:commutation_relation_theta_phi}\end{align}
The contribution in squared brackets in Eq. (\ref{eq:density_bosonized})
describe long-wavelength density fluctuations around the average value
$\rho_{0}$, while in the last term, each contribution describes low-energy
density fluctuations of momentum $k\sim2p\rho_{0}$, where $p$ is
an integer. When replaced into the effective action Eq. (\ref{eq:S_2nd_order})
we obtain the following effective model\begin{align}
S_{\text{eff}}\left[\phi\right] & =S_{0}\left[\phi\right]+S_{1}\left[\phi\right]+S_{2}\left[\phi\right],\label{eq:S_eff}\end{align}
where the contribution $S_{0}$ corresponds to a Luttinger liquid
model\citep{giamarchi_book_1d} \begin{align}
S_{0}\left[\phi\right] & =\frac{1}{2\pi\beta L}\sum_{k,\omega_{m}}\left[\frac{\omega_{m}^{2}}{uK}+\frac{uk^{2}}{K}+\frac{\eta\left|\omega_{m}\right|\left|k\right|}{2\pi c}\right]\left|\phi\left(k,\omega_{m}\right)\right|^{2}.\label{eq:S_0}\end{align}
resulting from the slow fluctuations of the density $\delta\rho\left(x\right)\sim-\nabla\phi\left(x\right)/\pi$
and from the hydrodynamic (i.e., $\left\{ k,\omega_{m}\right\} \rightarrow0$)
sector of $v_{\textrm{eff}}\left(k,\omega_{m}\right)$. $u$ and $K$
are respectively the velocity of the 1D plasmon and the interaction
Luttinger parameter

\begin{align}
K & \equiv\pi\sqrt{\frac{E_{J}}{E_{0}}},\label{eq:parameter_K}\\
u & \equiv a\sqrt{E_{J}E_{0}},\label{eq:parameter_u}\end{align}
where $E_{0}\equiv e^{2}/2C_{0}$ is the charging energy with respect
to the ground, with $C_{0}=\epsilon_{\text{r}}a/4\ln\left(2d/a\right)$
the effective ground capacitance of the Josephson junction. In our
treatment, due to the screening provided by the 2DEG, the static effective
potential $v_{\textrm{eff}}\left(k,0\right)$ is effectively short-ranged
for distances $x\gg d$, and therefore the Luttinger parameter $K$
is a constant.\citep{giamarchi_book_1d} %
\footnote{Indeed, in the static limit $\omega_{m}=0$, the effective screening
potential writes $v_{\textrm{scr}}\left(k,0\right)=2e^{2}K_{0}\left(2kd\right)/\epsilon_{\text{r}}$
and in the limit $kd\rightarrow0$ compensates the logarithmic divergence
of $v_{\textrm{1D}}\left(k,0\right)$ {[}cf. Eq. (\ref{eq:v_eff}){]},
and yields $\lim_{k\rightarrow0}v_{\textrm{eff}}\left(k,0\right)=2e^{2}\ln\left(2d/a\right)/\epsilon_{\text{r}}$.\citep{Lobos10_Dissipative_phase_fluctuations} %
}In terms of the capacitance matrix $C_{ij}$ of the 1DJJA, this amounts
to neglecting the interjunction capacitance $C$, since this contribution,
although relevant for density fluctuations of momentum $k\sim a^{-1}$,
drops off in the long-wavelength sector $k\rightarrow0$ (i.e. the
interaction is screened in a length $L_{\text{scr}}\sim a\sqrt{C/C_{0}}$).\citep{fazio01_review_superconducting_networks}
In the present context, the Luttinger parameter $K$ physically represents
the competition between coherence and charging effects in the array
{[}cf. Eq. (\ref{eq:parameter_K}){]}. Therefore, a large parameter
$K$ favors superconducting correlations, while a small value of $K$
tends to destroy superconductivity due to strong charging effects.\citep{fazio01_review_superconducting_networks} 

The dissipative parameter $\eta$ is defined as\begin{align}
\eta & \equiv\frac{c}{\epsilon_{\text{r}}8\pi}\frac{R_{\square}}{R_{Q}},\label{eq:eta_parameter}\end{align}
where $R_{\square}$ is the sheet-resistance of the 2D film and $c$
is a numerical constant of order $c\sim\mathcal{O}\left(1\right)$.
Eq. (\ref{eq:S_0}) with a non-vanishing $\eta$ describes a 1D plasmon-mode
with a finite lifetime $\Gamma\sim\left|k\right|/\eta$.\citep{Lobos10_Dissipative_phase_fluctuations}
Physically, a broadening of the 1D plasma mode occurs due to coupling
to the diffusive modes in the 2DEG. The term $\sim\eta\left|\omega_{m}\right|\left|k\right|$
in Eq. (\ref{eq:S_0}) is the result of combining the leading contribution
in powers of $\left|\omega_{m}\right|/Dk_{\text{TF}}\left|k\right|$
(with $k_{\text{TF}}$ the Thomas-Fermi momentum in the 2DEG) in the
expansion of the retarded potential $v_{\text{eff}}\left(k,\omega_{m}\right)$,
and the long-wavelength fluctuations of the density $\sim\left(\nabla\phi\right)^{2}$,
which contributes a term $\sim k^{2}\left|\phi\left(k,\omega_{m}\right)\right|^{2}$
in Fourier representation. Note that since the scaling dimension of
the term $\sim\left|k\right|\left|\omega_{m}\right|$ is $2$, the
critical properties of the system are not modified. Moreover, for
a metallic plane {[}cf. Ref. \onlinecite{rimberg97_dissipation_driven_sit_2D_josephson_array}{]}
with $R_{\square}\sim0.1R_{Q}$, $\eta\simeq10^{-2}\ll1$ and it can
be effectively ignored, allowing us to write\begin{align}
S_{0}\left[\phi\right] & \simeq\frac{1}{2\pi\beta L}\sum_{k,\omega_{m}}\left[\frac{1}{uK}\omega_{m}^{2}+\frac{u}{K}k^{2}\right]\left|\phi\left(k,\omega_{m}\right)\right|^{2}.\label{eq:S_0-1}\end{align}
This assumption greatly simplifies the analysis, since the action
$S_{0}$ recovers Lorentz-invariance in space-time.

The next term $S_{1}$ in Eq. (\ref{eq:S_eff}) originates in the
phenomenological potential $V_{l}\left(x\right)$, which has the same
periodicity of the array. Therefore, it can be decomposed in Fourier
components as $V_{l}\left(x\right)=\sum_{n}V_{n}\cos\left(Qnx\right),$
with $Q=2\pi/a$. In general, all terms other than $p=n=0$ in Eq.
(\ref{eq:density_bosonized}) are rapidly oscillating and vanish under
the integral sign. However, if $Qn=2\pi p\rho_{0}$, or equivalently
$\rho_{0}a=n/p$ (i.e., the average density of bosons is commensurate
with the lattice), then the term $\sim\int dx\; V_{l}\left(x\right)\rho\left(x\right)$
yields a term $e^{-i\left(Qn-2\pi p\rho_{0}\right)x}=1$ which is
not oscillating and, in addition to the term $p=n=0$, we have the
additional term \begin{align}
S_{1}\left[\phi\right] & =-\frac{\lambda}{a\tau_{0}}\int dxd\tau\;\cos\left(2\phi\left(x,\tau\right)\right),\label{eq:S_1}\end{align}
where we have only kept the most important commensurability ($p=1$),
and where we have defined the dimensionless parameter $\lambda$ \begin{align}
\lambda & \equiv V_{1}\tau_{0}.\label{eq:lambda_parameter}\end{align}
%
\begin{comment}
Note that $\lambda$ is dimensionless. Then the phase-slips rate is
\begin{align*}
\Gamma & \equiv\frac{\lambda}{\tau_{0}}=\frac{a\rho_{0}V_{1}}{\hbar}.\end{align*}
Then \begin{align*}
\lambda & =\Gamma\tau_{0}\\
 & =\frac{\Gamma a}{u}\\
 & =\frac{\hbar\Gamma}{\sqrt{E_{J}E_{C}}}\\
 & =\frac{10^{-34}J.s\times10^{9}s^{-1}}{K.k_{B}\sqrt{3.3\times0.17}}\\
 & =\frac{10^{-34}J.s\times10^{9}s^{-1}}{K.1.38\times10^{-23}J.K^{-1}\sqrt{3.3\times0.17}}\\
 & =\frac{10^{-34}\times10^{9}}{1.38\times10^{-23}\sqrt{3.3\times0.17}}\\
 & \simeq10^{-34+23+9}\sim0.01\end{align*}

\end{comment}
{}and the short-time cutoff $\tau_{0}\equiv a/u$. The term $V_{0}$
can be reabsorbed in a redefinition of the chemical potential of the
externally imposed charge $\bar{n}$, so in the following we will
not consider it. Physically, the dimensionless parameter $\lambda$
is related to the QPS rate in the Josephson junction by $\Gamma_{\text{QPS}}=\lambda/\tau_{0}$.
Estimated experimental values for $\Gamma_{\text{QPS}}$ are in the
order of $\sim1\;\text{GHz}$\citep{Manucharyan10_Coherent_QPS_in_JJA},
which yields $\lambda\simeq0.06$.

The final term in Eq. (\ref{eq:S_eff}) comes from the dissipative
part of $v_{\textrm{scr}}\left(k,\omega_{m}\right)$. Due to the strongly
oscillating factors $\sim e^{-i2\pi p\rho_{0}x}$ in Eq. (\ref{eq:density_bosonized}),
it results in the local dissipative term \begin{align}
S_{2}\left[\phi\right] & =-\frac{\eta}{a}\int dxd\tau d\tau^{\prime}\;\sum_{p>0}\frac{1}{p}\frac{\cos2p\left[\phi\left(x,\tau\right)-\phi\left(x,\tau^{\prime}\right)\right]}{\left(\tau-\tau^{\prime}\right)^{2}},\label{eq:S_2}\end{align}
This contribution is consistent with that of Ref. \onlinecite{cazalilla06_dissipative_transition},
obtained in the context of Luttinger liquids capacitively coupled
to diffusive metals. In spite of the small magnitude of $\eta$, we
will show that this contribution has important consequences for the
critical properties of the 1DJJA, in contrast to the term proportional
to $\eta$ in Eq. (\ref{eq:S_0}).

In the following we study the critical properties and phases of the
model obtained in Eq. (\ref{eq:S_eff}).

\section{\label{sec:phase-diagram}Phase diagram}

\subsection{\label{sub:RG}Weak-coupling renormalization group analysis}

We first focus on the phases of the 1DJJA at $T=0$. To that end,
we perform a weak-coupling renormalization group (RG) analysis of
the model Eq. (\ref{eq:S_eff}), assuming that $S_{1}$ and $S_{2}$
in Eqs. (\ref{eq:S_1}) and (\ref{eq:S_2}) respectively are weak
perturbations to the Luttinger liquid $S_{0}$ in Eq. (\ref{eq:S_0-1}).
Since the action $S_{0}$ is Lorentz-invariant in space and imaginary
time, we adopt an RG procedure that rescales homogenously space and
time. As usual, we assumme that the original theory is defined up
to a certain momentum cutoff $\Lambda\left(l\right)=\Lambda_{0}e^{-l}$
(with $\Lambda_{0}\sim a^{-1}$), and we study how the action $S_{0}$
is renormalized upon integration of high-energy modes in a window
between $\Lambda\left(l\right)/s<\left|\begin{gathered}\mathbf{q}\end{gathered}
\right|<\Lambda\left(l\right)$, with $s=e^{dl}$, where we have employed the compact notation $\mathbf{q}\equiv\left\{ k,-\frac{\omega_{m}}{u}\right\} $
and $\mathbf{x}\equiv\left\{ x,u\tau\right\} $. 

We obtain the perturbative RG-flow equations of the model by performing
a one-loop correction in $S_{2}$ and a two-loop correction in $S_{1}$,
and requiring that the term $S_{0}$ is invariant upon scaling.\citep{Shankar}
We obtain the RG-flow equations

\begin{align}
\frac{dK\left(l\right)}{dl} & =\left[-2\pi\eta\left(l\right)-\left(2\pi\right)^{2}K\left(l\right)\lambda^{2}\left(l\right)C\right]K^{2}\left(l\right),\label{eq:RG_eq_K}\\
\frac{du\left(l\right)}{dl} & =-2\pi\eta\left(l\right)u\left(l\right)K\left(l\right),\label{eq:RG_eq_u}\\
\frac{d\lambda\left(l\right)}{dl} & =\left[2-K\left(l\right)\right]\lambda\left(l\right),\label{eq:RG_eq_lambda}\\
\frac{d\eta\left(l\right)}{dl} & =\left[1-2K\left(l\right)\right]\eta\left(l\right),\label{eq:RG_eq_eta}\end{align}
where the numerical constant $C$ is of order unity. 

Note that both $S_{1}$ and $S_{2}$ tend to destroy superconducting
correlations in the Luttinger liquid phase, a fact that is reflected
in Eq. (\ref{eq:RG_eq_K}) where the Luttinger parameter $K\left(l\right)$
is renormalized to \textit{smaller} values, meaning that charging
effects are enhanced. This can be interpreted as an effective increase
of the charging energy $E_{0}$ in Eq. (\ref{eq:parameter_K}). In
addition, since $S_{2}$ is the only term that breaks the Lorentz
invariance of the theory, note that the plasmon velocity $u\left(l\right)$
is proportional only to $\eta\left(l\right)$, and is independent
of $\lambda\left(l\right)$.

\begin{figure}[h]
\includegraphics[clip,scale=0.9]{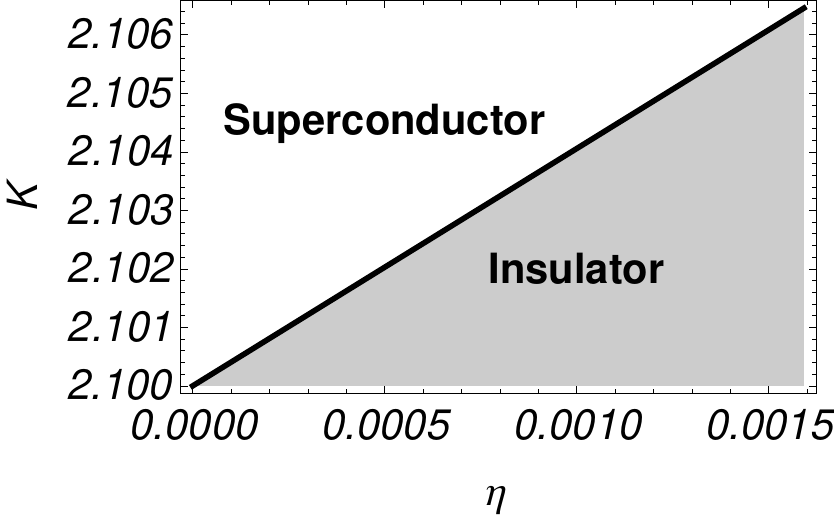}\caption{Schematic phase diagram of the 1DJJA in the $K-\eta$ plane, obtained
from the integration of the RG-flow Eqs. (\ref{eq:RG_eq_K})-(\ref{eq:RG_eq_eta}),
with the initial parameter $\lambda_{0}\equiv\lambda\left(l=0\right)=0.01$.
An increase of $R_{\square}$ in the 2DEG, and consequently, of the
dissipative parameter $\eta$, can induce a SIT. Note that, in absence
of dissipation, the critical value $K_{c}=K_{c}\left(\lambda_{0}\right)\simeq2.1$
is slightly shifted with respect to the value $K_{c}\left(\lambda_{0}\rightarrow0\right)\rightarrow2$.\label{fig:phase_diagram}}

\end{figure}

When $K\left(l\right)<2$, the perturbative parameter $\lambda\left(l\right)$
flows to strong-coupling {[}cf. Eq. (\ref{eq:RG_eq_lambda}){]}, and
the perturbative RG procedure is no longer valid. In the limit $\eta\rightarrow0$
we recover the usual Mott-transition of the BKT-type described by
the sine-Gordon model, and below the critical value $K_{c}=2$, the
1DJJA is in the insulating phase.\citep{bradley_josephson_chain,glazman_josephson_1d,fazio01_review_superconducting_networks,giamarchi_book_1d}
Using Eq. (\ref{eq:parameter_K}), this means that in absence of dissipation,
the SIT occurs for $E_{J}/E_{0}=\left(2/\pi\right)^{2}$.\citep{fazio01_review_superconducting_networks}
Note that our situation corresponds strictly to the case when the
superfluid density in the 1DJJA is commensurate to the lattice, and
is in clear distinction to the non-commensurate situation (i.e., $\lambda=0$),
where dissipation (i.e., the term $S_{2}$) becomes relevant for $K\left(l\right)<1/2$,
inducing a different kind of non-superconducting groundstate.\citep{cazalilla06_dissipative_transition}

In the present case, the scaling dimension of the dissipative parameter
$\eta\left(l\right)$ is always smaller than that of $\lambda\left(l\right)$,
which means that for $K\left(l\right)\simeq2$, $S_{1}$ is a stronger
perturbation as compared to $S_{2}$. Therefore, one would expect
the nature of the non-superconducting groundstate to be determined
essentially by $S_{1}$. However, based on this fact, one could naively
conclude that the term $S_{2}$ is unimportant near the SIT, a conclusion
we prove incorrect. In fact, a more detailed analysis reveals the
importance of the term $S_{2}$ near the SIT. Physically, the coupling
to the diffusive degrees of freedom in the 2DEG quenches charge-fluctuations
in the 1DJJA, resulting in an enhanced effective charging energy $E_{0}^{*}$.
This phenomenon is more precisely described by the RG-flow equation
for $K\left(l\right)$ {[}cf. Eq. (\ref{eq:RG_eq_K}){]}, where $K\left(l\right)$
is renormalized to \textit{lower} values by $\eta\left(l\right)$.
Indeed, near the SIT, a small increase in the initial value $\eta_{0}\equiv\eta\left(l=0\right)$
(i.e., an increase in $R_{\square}$) can effectively control the
RG-flow of $K\left(l\right)$ and therefore, that of $\lambda\left(l\right)$,
inducing the SIT. We illustrate this point in Fig \ref{fig:phase_diagram},
where the schematic phase diagram obtained by integration of the RG-flow
Eqs. (\ref{eq:RG_eq_K})-(\ref{eq:RG_eq_eta}), with initial parameter
$\lambda_{0}\equiv\lambda\left(l=0\right)=0.01$. Note the stabilization
of the insulating groundstate due to Ohmic dissipation induced by
the coupling to the 2DEG. 

In a first approximation, this effect is similar to the dissipation-driven
SIT observed in 2DJJAs capacitively coupled to a diffusive 2DEG.\citep{rimberg97_dissipation_driven_sit_2D_josephson_array,Wagenblast97_SIT_in_a_Tunable_Dissipative_Environment,Vishwanath04_Screening_and_dissipation_at_the_SIT_induced_by_a_metallic_ground_plane}
However, important differences appear with respect to the 2D case.
In that case, it was argued that dissipation produced a renormalization
of the effective parameters $E_{J}$ and $E_{C}$ of the array due
to the incomplete screening of the Coulomb interaction in a certain
frequency-regime.\citep{Wagenblast97_SIT_in_a_Tunable_Dissipative_Environment}
Physically, the slow diffusive response of the 2DEG cannot follow
the faster dynamics of the 2D plasma mode, and cannot screen it efficiently.
However, in the 1D geometry the 1D plasmon is effectively very well
screened by the 2DEG\citep{Lobos10_Dissipative_phase_fluctuations},
and it could be naively concluded that no dissipation-driven SIT should
be observed. However, this screening effect is compensated by the
presence of strong backscattering occuring in 1D {[}i.e., action $S_{2}$,
Eq. (\ref{eq:S_2}){]}, and originated in the retarded interaction
$v_{\text{eff}}\left(x,\tau\right)$. The net result is that the dissipation-driven
SIT is restored in 1D. 

Although one expects the nature of the non-superconducting groundstate
to be of the Mott-insulating type, by analogy with the well-known
results for the sine-Gordon model\citep{bradley_josephson_chain,glazman_josephson_1d,fazio01_review_superconducting_networks,giamarchi_book_1d},
strictly speaking we cannot extrapolate the results in this Section
to the strong-coupling situation, and a different method is needed
in that regime.

\subsection{\label{sub:SCHA}Self-consistent harmonic approximation}

To gain more insight into the phase in which the parameter $\lambda\left(l\right)$
flows to strong-coupling, in this Section we make use of the variational
self-consistent harmonic approximation\citep{feynman_statmech}. This
method consist in finding the optimal propagator $g_{\text{tr}}^{-1}\left(\mathbf{q}\right)$
of a Gaussian trial action of the 1DJJA

\begin{align}
S_{\text{tr}} & =\frac{1}{2\beta L}\sum_{\mathbf{q}}g_{\text{tr}}^{-1}\left(\mathbf{q}\right)\left|\phi_{\mathbf{q}}\right|^{2},\label{eq:S_tr}\end{align}
where $\phi_{\mathbf{q}}$ is the Fourier transform of $\phi\left(x,\tau\right)$.
Here we have introduced the compact notation $\mathbf{q}=\left(k,-\omega_{m}/u\right)$.
The idea is to minimize the variational free-energy

\begin{eqnarray}
F_{\text{var}} & \equiv & F_{\text{tr}}+\frac{1}{\beta}\left\langle S_{\text{eff}}-S_{\text{tr}}\right\rangle _{\text{tr}},\label{eq:F_var}\end{eqnarray}
where the {}``trial'' free-energy $F_{\text{tr}}$ is \begin{align}
F_{\text{tr}} & =-\frac{1}{2\beta}\sum_{\mathbf{q}}\log\left[\beta Lg_{\text{tr}}\left(\mathbf{q}\right)\right]\label{eq:F_tr}\end{align}
The factor $1/2$ in Eqs. (\ref{eq:S_tr}) and (\ref{eq:F_tr}) come
from the constraint $\phi^{*}\left(\mathbf{q}\right)=\phi\left(-\mathbf{q}\right)$
since $\phi\left(x,\tau\right)$ is a real field, a fact that reduces
the number of independent degrees of freedom . 

\begin{figure}[h]
\includegraphics[bb=40bp 85bp 420bp 340bp,clip,scale=0.6]{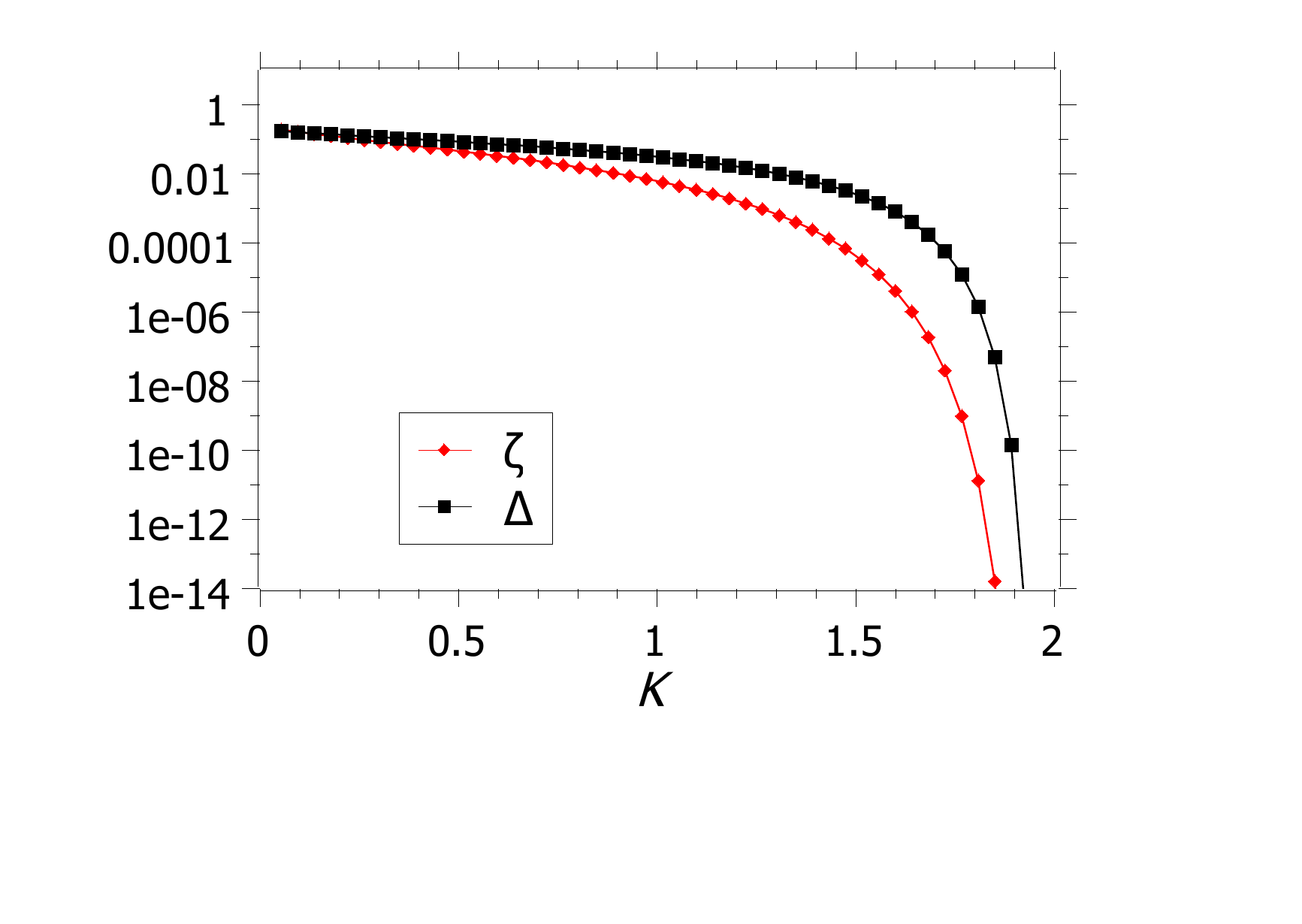}\caption{Dimensionless SCHA parameters $\Delta$ and $\zeta$ as a function
of $K$, calculated for the parameters $\eta=0.01$, $\lambda=0.05$.
We obtain non-vanishing values only in the region $K\lesssim2$ (note
the abrupt increase in that region), consistent with the results of
the RG-analysis. \label{fig:scha}}

\end{figure}

The minimization of $F_{\text{var}}$ in Eq. (\ref{eq:F_var}) with
respect to $g_{\text{tr}}\left(\mathbf{q}\right)$ yields the self-consistent
equation for $g_{\text{tr}}\left(\mathbf{q}\right)$\begin{align}
g_{\text{tr}}^{-1}\left(\mathbf{q}\right) & =\frac{1}{\pi uK}\omega_{m}^{2}+\frac{u}{\pi K}k^{2}+\frac{4\lambda}{a\tau_{0}}e^{-\frac{2}{\beta L}\sum_{\mathbf{q}^{\prime}}g_{\text{tr}}\left(\mathbf{q}^{\prime}\right)}\nonumber \\
 & -\frac{8\eta}{a}\int_{0}^{\beta}d\tau\;\frac{\cos\left(\omega_{m}\tau\right)-1}{\tau^{2}}\times\nonumber \\
 & \times e^{-4\frac{1}{\beta L}\sum_{\mathbf{q}^{\prime}}\left[1-\cos\left(\omega_{m}^{\prime}\tau\right)\right]g_{\text{tr}}\left(\mathbf{q}^{\prime}\right)}.\label{eq:self_consistent_scha}\end{align}
In general, the solution of this equation has to be found numerically.
However, for small $\lambda$ and $\eta$ the analytical solution
\begin{align}
g_{\text{tr}}^{-1}\left(\mathbf{q}\right) & =g_{LL}^{-1}\left(\mathbf{q}\right)+\frac{\zeta}{a}\left|\omega_{m}\right|+\frac{\Delta}{a\tau_{0}},\label{eq:g0_combined}\end{align}
is obtained. Here $g_{LL}^{-1}\left(\mathbf{q}\right)=\frac{1}{\pi uK}\omega_{m}^{2}+\frac{u}{\pi K}k^{2}$
is the Luttinger liquid propagator, corresponding to the action Eq.
(\ref{eq:S_0}) (for $\eta=0$). Physically, this propagator describes
an insulator (given by a non-vanishing gap or {}``mass'' term $\Delta$)
with Ohmic-dissipative dynamics (encoded in a non-vanishing $\zeta$).
Note that dissipation dominantes for frequencies $\left|\omega_{m}\right|>\Delta/\zeta\tau_{0}$.
These parameters are found solving the following set of non-linear
equations \begin{align}
\zeta & =8\pi\eta\left(\frac{\zeta K\pi+2\sqrt{K\pi\Delta}}{4}\right)^{2K},\label{eq:scha_zeta}\\
\Delta & =4\lambda\left(\frac{\zeta K\pi+2\sqrt{K\pi\Delta}}{4}\right)^{K},\label{eq:scha_Delta}\end{align}
obtained replacing the solution Eq. (\ref{eq:g0_combined}) back into
Eq. (\ref{eq:self_consistent_scha}). Starting from the self-consistent
solution of Eqs. (\ref{eq:scha_zeta}) and (\ref{eq:scha_Delta})
for $\Delta$ in absence of dissipation (i.e. $\eta=0$), we can study
the regime $\eta\ll\lambda\ll1$ perturbatively in $\eta$, and we
obtain the following estimate for the gap increase due to dissipative
effects \begin{align}
\delta\Delta & \simeq2\pi^{2}\frac{\eta K\Delta_{0}^{2}}{\lambda}.\label{eq:gap_increase}\end{align}
This result is consistent with the fact that dissipation in the density
(i.e., field $\phi$) quenches charge-fluctuations and therefore favors
an insulating groundstate. 

In Fig. \ref{fig:scha} we show numerical results for $\Delta$ and
$\zeta$ as a function of $K$ for the values $\lambda=0.05$ and
$\eta=0.0\text{1}$. Note the sharp increase of both $\Delta$ and
$\zeta$ for $K<2$. This result is consistent with the RG-analysis,
which predict the breakdown of the Luttinger liquid phase for $K<2$
in the weak-coupling regime. Within the SCHA, the physics of the strong-coupling
fixed point is encoded in non-vanishing values of $\zeta$ and $\Delta$,
providing a complementary description to the RG-analysis.

\section{\label{sec:transport}Transport properties}

In this section we concentrate on the dc-conductivity of the 1DJJA,
a quantity of central interest in experiments.\citep{schoen90_review_ultrasmall_tunnel_junctions,fazio01_review_superconducting_networks}
We first focus on the current-density $j\left(x\right)$. Since the
field $\nabla\theta\left(x\right)/\pi$ is the momentum of Cooper-pairs
{[}cf. Eq. (\ref{eq:H_JJA_continuum}){]}, the usual minimal coupling
procedure $\nabla\theta\left(x\right)/\pi\rightarrow\left[\nabla\theta\left(x\right)-2eA\left(x\right)\right]/\pi$
(with $e$ the electron charge and $A$ the vector potential) in Hamiltonian
Eq. (\ref{eq:H_JJA_continuum}) allows to obtain the current as $j\left(x\right)\equiv-\delta H_{JJA}/\delta A\left(x\right)$.
In our problem, it explicitly reads \citep{giamarchi_book_1d} \begin{align}
j\left(x\right) & =uK\left(\frac{2e}{\pi}\right)\left[\nabla\theta\left(x\right)-2eA\left(x\right)\right].\label{eq:current_density}\end{align}
The conductivity along the wire is obtained from the Kubo formula
\citep{mahan2000,giamarchi_book_1d} \begin{align}
\sigma\left(\omega\right) & \equiv\frac{\chi_{jj}^{R}\left(0,\omega\right)}{i\left(\omega+i\delta\right)},\label{eq:conductivity}\end{align}
where $\chi_{jj}^{R}\left(k,\omega\right)\equiv\lim_{i\omega_{m}\rightarrow\omega+i\delta}\chi_{jj}\left(\mathbf{q}\right)$
is the retarded current-current correlation function and $\chi_{jj}\left(\mathbf{q}\right)\equiv\left\langle j^{*}\left(\mathbf{q}\right)j\left(\mathbf{q}\right)\right\rangle =\left.\delta^{2}\ln Z/\delta A\left(\mathbf{q}\right)\delta A^{*}\left(\mathbf{q}\right)\right|_{A=0}$
is the current-current correlation function obtained in the linear-response
regime. It is convenient to express this correlator as $\chi_{jj}\left(\mathbf{q}\right)=\chi_{jj}^{d}+\chi_{jj}^{p}\left(\mathbf{q}\right)$,
where $\chi_{jj}^{d}\equiv-\left(2e\right)^{2}uK/\pi$ is the diamagnetic
contribution and 

\begin{align}
\chi_{jj}^{p}\left(\mathbf{q}\right) & \equiv\left(\frac{2e}{\pi}\right)^{2}\left(uK\right)^{2}k^{2}\left\langle \theta\left(\mathbf{q}\right)\theta\left(-\mathbf{q}\right)\right\rangle ,\label{eq:xi_p}\end{align}
is the paramagnetic term.\citep{mahan2000} In absence of current-decaying
mechanisms {[}i.e., $\lambda=\eta=0$ in Eq. (\ref{eq:S_eff_total}){]},
the conductivity writes\begin{align*}
\sigma_{0}\left(\omega\right) & =\frac{\left(2e\right)^{2}}{\hbar}uK\left[\delta\left(\omega\right)+i\text{P}\left(\frac{1}{\pi\omega}\right)\right],\end{align*}
where we have restored the Planck constant and where we have used
that $\chi_{jj}^{p}\left(\mathbf{q}\right)\rightarrow0$ in the limit
$k=0$.\citep{giamarchi_book_1d} Note that the real part of $\sigma_{0}\left(\omega\right)$
consists of a Drude-peak at $\omega=0$, as expected for a superconductor.
This result can be understood from the fact that the total charge
current $J_{e}=\int dx\; j\left(x\right)$ is a conserved quantity
in absence of QPS and dissipation processes, i.e., it commutes with
the hamiltonian $H_{\text{JJA}}$.

The effect of a finite $\eta$ in the Gaussian sector of the theory
{[}cf. Eq. (\ref{eq:S_0}){]} has been studied in Ref. \onlinecite{Lobos10_Dissipative_phase_fluctuations},
and produces a broadening of the plasmon peak, whose width $\Gamma$
vanishes as $\Gamma\sim\left|k\right|$. Consequently, only taking
into account this effect, a well-defined Drude-peak in $\sigma\left(\omega\right)$
for $\omega=0$ is recovered, and the system should behave as a perfect
conductor. 

Let us now study the effects of the terms $S_{1}$ and $S_{2}$. When
$\lambda$ and $\eta$ are irrelevant perturbations (in the RG sense),
their effects on the conductivity can be studied within the theoretical
framework of the memory function formalism.\citep{gotze_fonction_memoire}
In this approach, the central assumption is that the Kubo formula
for the conductivity Eq. (\ref{eq:conductivity}) can be recasted
as\citep{giamarchi_book_1d} \begin{align}
\sigma\left(\omega,T\right) & =\frac{i\left(2e\right)^{2}}{\pi\hbar}\frac{uK}{\omega+M\left(\omega,T\right)},\label{eq:sigma_omega}\end{align}
where $M\left(\omega,T\right)$ (i.e., the memory function) is a meromorphic
function depending on the terms in the Hamiltonian responsible for
degrading the current, and hence producing a finite resistivity. Current-decay
originated in QPS and in the coupling to the dissipative modes in
the 2DEG induce finite resistivity in the 1DJJA for all temperatures
$T<T_{c}$. In particular for temperatures $T\ll T_{c}$, and perturbatively
in $\lambda$ and $\eta$, we obtain\begin{align}
\varrho\left(T\right) & =\frac{\hbar}{a\left(2e\right)^{2}}\left[A_{1}T^{2K-3}+A_{2}T^{2K}\right]\label{eq:resistivity}\end{align}
where 

\begin{align}
A_{1} & \equiv\lambda^{2}4\pi^{3}\left[\cos\left(\frac{\pi K}{2}\right)B\left(\frac{K}{2},1-K\right)\right]^{2}\left(\frac{2\pi a}{u}\right)^{2K-3},\label{eq:A_1}\\
A_{2} & \equiv\eta32\pi^{3}\cos\left[\left(1+K\right)\pi\right]B\left[1+K,-1-2K\right]\left(\frac{2\pi a}{u}\right)^{2K},\label{eq:A_2}\end{align}
where the function $B\left(x,y\right)$ is defined as $B\left(x,y\right)\equiv\Gamma\left(x\right)\Gamma\left(y\right)/\Gamma\left(x+y\right)$,
and $\Gamma\left(x\right)$ is the standard Euler's Gamma function.\citep{abramowitz_math_functions}
The term $\sim T^{2K-3}$ in Eq. (\ref{eq:resistivity}) is the contribution
due to QPS processes, consistent with former theoretical predictions.\citep{giamarchi_attract_1d,zaikin97}
The second term $\sim T^{2K}$ originates in backscattering effects
induced by dissipation, and is consistent with the behavior predicted
by Cazalilla \textit{et al.}\citep{cazalilla06_dissipative_transition}
This last effect can be interpreted as a frictional drag produced
by the diffusive modes in the 2DEG.\citep{rojo99_coulomb_drag_review}
Note that at lowest order in $\lambda$ and $\eta$, the two contributions
add up independently, indicating that for temperatures $T^{*}<T\ll T_{c}$,
where $T^{*}\equiv\sqrt[3]{A_{1}/A_{2}}/2\pi\tau_{0}$, the resistivity
in the 1DJJA is dominated by frictional drag, while for $T<T^{*}\ll T_{c}$
the effect of QPS takes over.

The non-trivial effects due to the renormalization of the bare couplings
can be taken into account integrating the RG-flow Eqs. (\ref{eq:RG_eq_K})-(\ref{eq:RG_eq_eta}),
and injecting them in the above Eqs. (\ref{eq:resistivity}), (\ref{eq:A_1})
and (\ref{eq:A_2}). We integrate the RG-flow up to the scale given
by the temperature $a\left(l\right)=a\left(0\right)e^{l}=u\left(l\right)/2\pi T$,
and we use formula Eq. (\ref{eq:resistivity}) with the parameters
of the model calculated at the scale $a\left(l\right)$. This allows
to obtain $\varrho\left(T\left(l\right)\right)$ vs $T\left(l\right)$.

\begin{figure}[h]
\includegraphics[bb=25bp 450bp 500bp 820bp,clip,scale=0.47]{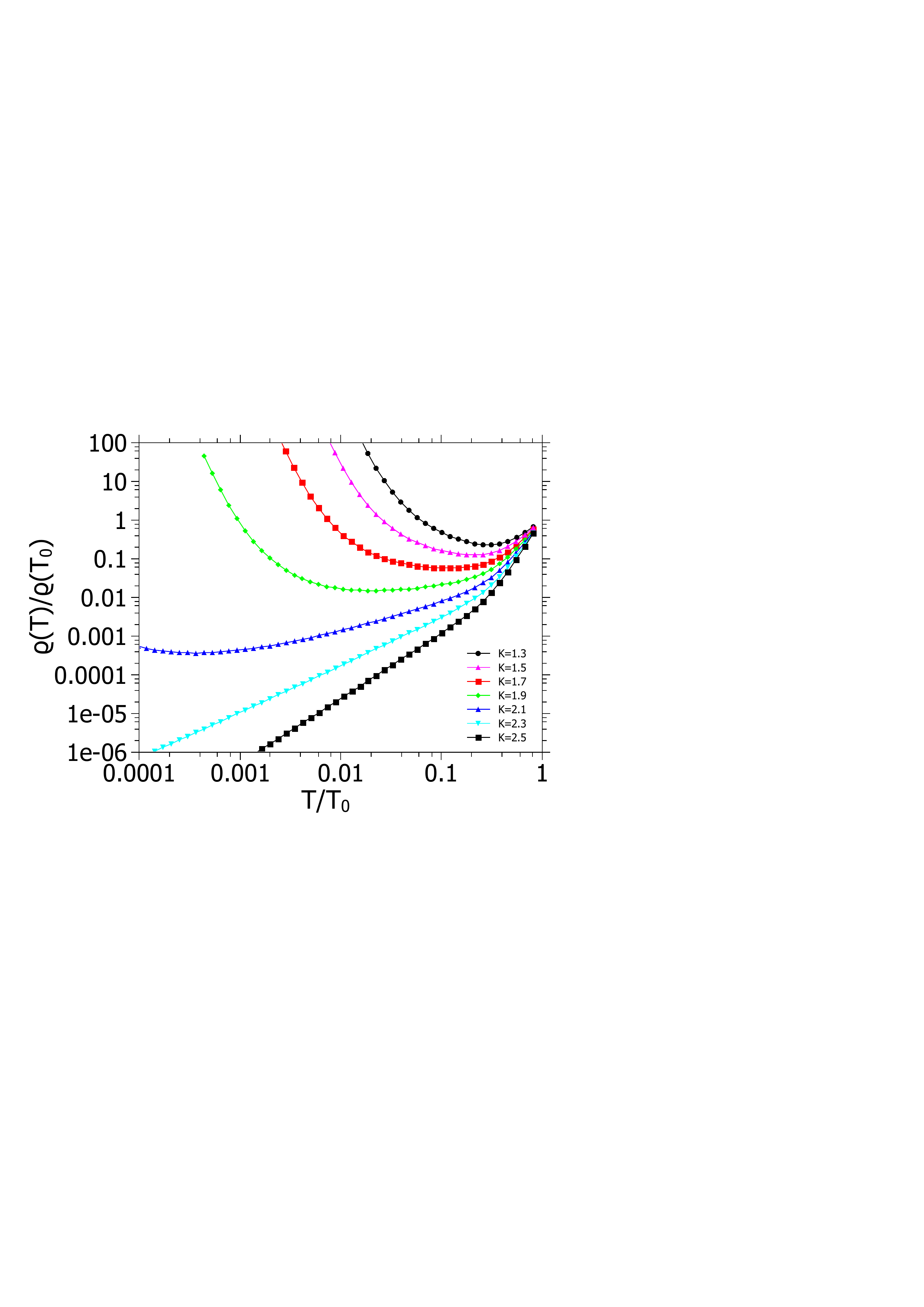}\caption{\label{fig:resistivity_log_log}Dc-resistivity $\varrho\left(T\right)$
of the 1DJJA, normalized to a {}``high-temperature'' value $\varrho\left(T_{0}\right)$,
as a function of $T/T_{0}$, calculated for the parameters $\lambda_{0}=0.01$
and $\eta_{0}=0.01$, and for different values of $K=\pi\sqrt{E_{J}/E_{0}}$.
A low-temperature upturn of $\varrho\left(T\right)$ signals the formation
of the insulating phase.}

\end{figure}

In Fig. \ref{fig:resistivity_log_log} we show the resistivity $\varrho\left(T\right)$
of the 1DJJA, calculated for different values of the parameter $K$
and using the estimations for the bare parameters $\lambda_{0}=0.01$
and $\eta_{0}=0.01$. The results are normalized to a {}``high-temperature''
resistivity $\varrho\left(T_{0}\right)$, where $T_{0}\simeq a\left(0\right)/u=\tau_{0}$,
represents a high-temperature cutoff in the theory (e.g., $T_{c}$). 

Note that for the values $K=2.5$ and $K=2.3$, the resistivity shows
a monotonically decreasing behavior, indicating a superconducting
groundstate and consistent with the RG-analysis of Sec. \ref{sub:RG}.
We also note a small kink around $T^{*}\sim0.4\; T_{0}$, signalling
the aforementioned crossover from dissipation-dominated to QPS-dominated
resistivity. For $K=2.1$, the resistivity first decreases and then
shows a low-temperature upturn, indicating that the array is near
the quantum critical point $K_{c}$. Finally, for lower values of
$K$, the insulating behavior in the 1DJJA is clear. Since both the
integration of the RG-flow equations and the calculation of the memory-function
formulas are perturbative in $\lambda$ and $\eta$, the calculation
of the resistivity must be stopped whenever $\lambda\left(l\right)$
or $\eta\left(l\right)$ become of order unity.

\begin{figure}[h]
\includegraphics[bb=30bp 30bp 450bp 350bp,clip,scale=0.55]{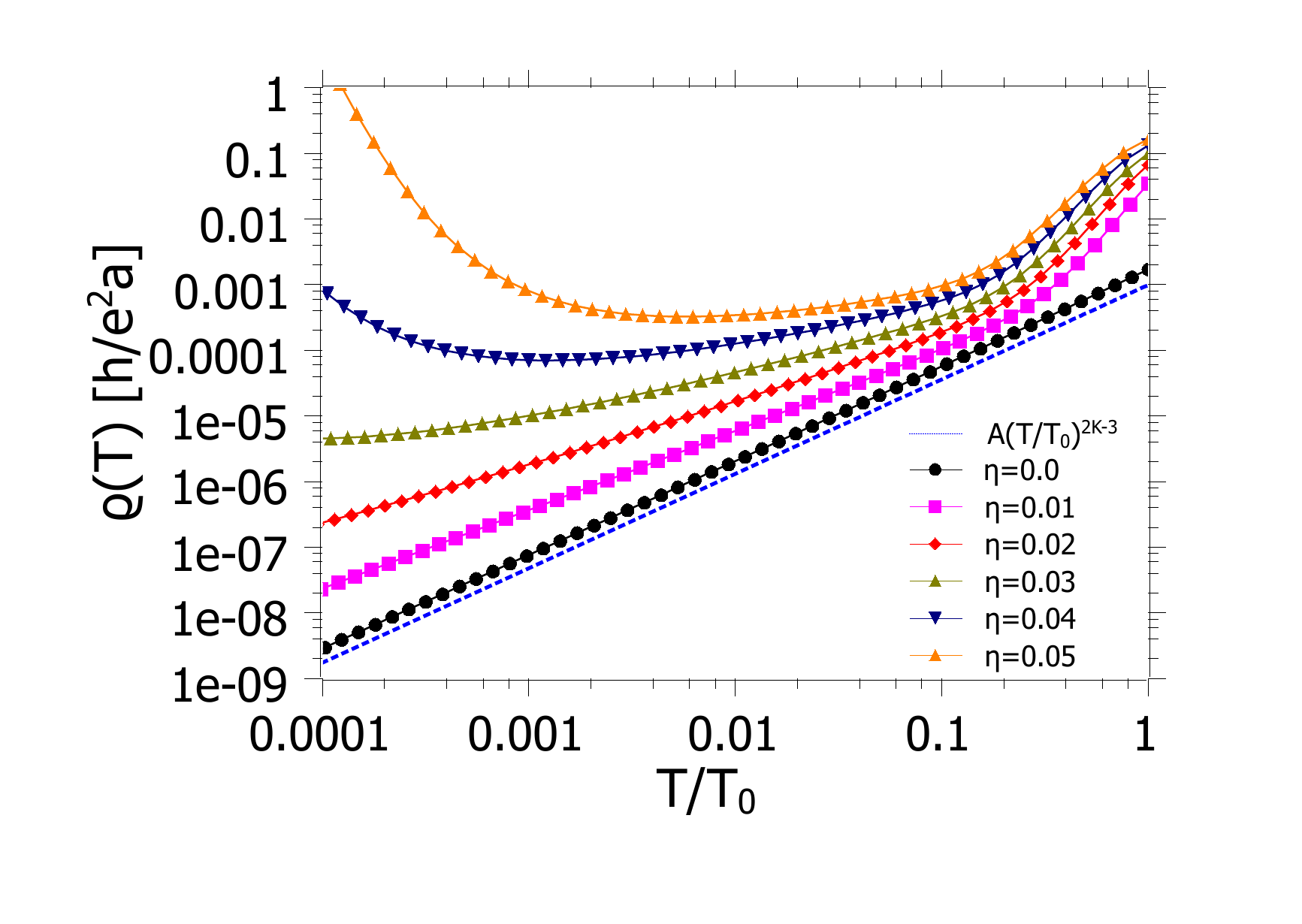}\caption{\label{fig:resistivity_eta}Resistivity of the 1DJJA (in units of
$h/e^{2}a$ ) as a function of $T/T_{0}$, calculated for parameters
$K=2.3$ and $\lambda=0.01$ and for different values of $\eta$.
Although in absence of dissipation, the array is in the superconducting
phase, a dissipation-driven SIT occurs upon increasing $\eta$, consistent
with the results in Fig. \ref{fig:phase_diagram}. The curve $\varrho\left(T\right)\sim T^{2K-3}$
is shown for comparison.}

\end{figure}

In Fig. \ref{fig:resistivity_eta}, we show the resistivity as a function
of $T/T_{0}$, calculated for fixed $K=2.3$ and $\lambda=0.01$,
and for different values of parameter $\eta$. We see that for $\eta=0$
(i.e., $R_{\square}=0$ in the 2DEG), the array shows superconducting
behavior, and the resistivity due to QPS processes is well described
by the predicted power-law $\varrho\left(T\right)\sim T^{2\tilde{K}-3}$,
with $\tilde{K}=K\left(l\rightarrow\infty\right)\simeq2.2$ the renormalized
value predicted by Eq. (\ref{eq:RG_eq_lambda}). Upon increasing the
parameter $\eta$, the resistivity of the array increases, developing
the aforementioned kink, but most importantly, the low-temperature
resistivity develops an upturn, indicating a dissipation-driven phase
transition to the insulating phase.

More insight into the insulating phase can be obtained using the Luther-Emery
refermionization solution for $K=1$.\citep{luther_emery_backscattering,giamarchi_book_1d}
In absence of dissipation (i.e., $\eta=0$) an exact solution is obtained
in terms of non-interacting fermions, with a gap $\tilde{\Delta}\equiv\pi a\lambda$
in their spectrum of excitations. Using the Kubo formula, one obtains
the following expression for the dc-conductivity at low temperatures
$T\ll\tilde{\Delta}$ 

\begin{align*}
\sigma\left(\omega\right) & \approx\frac{e^{2}}{\hbar}u\sqrt{\frac{2\pi T}{\tilde{\Delta}}}e^{-\tilde{\Delta}/T}\delta\left(\omega\right).\end{align*}
This contribution arises from the excited quasiparticles above the
gap $\tilde{\Delta}$, which have an exponential population at low
enough temperatures. This infinite conductivity occurs because in
absence of dissipation, excited quasiparticles are infinitely long-lived.
Using the memory-function approach for the refermionized problem in
the regime $\eta\ll\lambda,T\ll\tilde{\Delta}$, we find the analytical
result 

\begin{align*}
\sigma\left(\omega=0\right) & =\frac{e^{2}}{\hbar}\frac{c_{2}}{\eta}\frac{1}{\tau_{0}^{2}\tilde{\Delta}^{2}}\frac{\tilde{\Delta}}{T}e^{-\tilde{\Delta}/T},\end{align*}
where $c_{2}$ is a numerical coefficient $c_{2}\simeq\mathcal{O}\left(1\right)$.
As expected, dissipation introduced a finite lifetime in the quasiparticles,
and a finite resistivity is obtained at $\omega=0$.

\section{\label{sec:summary}Summary and conclusions}

We have investigated the properties of a linear JJA capacitively coupled
to a diffusive 2DEG placed in close proximity. Using a bosonization
approach, we have derived an effective model for the 1DJJA, and have
obtained its critical properties and phases at $T=0$. Our main result
is the possibility to observe a SIT tuned by the parameter $\eta\sim R_{\square}/R_{Q}$
{[}cf. Eq. (\ref{eq:eta_parameter}){]}. This setup could be used
to investigate the superconductor-insulator transition in a 1DJJA
under better controlled experimental conditions as compared to other
setups used in the past.\citep{Chow98_Length_scale_dependence_of_SIT_in_1D_array_J,Kuo01_Magnetic_induced_transition_in1DJJA,Miyazaki02_QPT_in_1D_arrays_of_JJs,Takahide06_SIT_2D_1D_crossover}
Our work could shed some light on the understanding of other 1D superconducting
systems showing a similar behavior, such as ultra-thin superconducting
wires built by molecular templating\citep{bezryadin00,lau01,Bezryadin08_QPS_review}
or by e-beam lithography\citep{Chen09_Magnetic-Field-Induced_Superconducting_State_in_Zn_Nanowires_Driven_in_the_Normal_State_by_an_Electric_Current}
techniques. 

We have shown that besides the more or less trivial static screening
effect, the presence of a 2DEG induces dissipative effects in the
quantum dynamics of the 1DJJA due to backscattering processes induced
by the dynamically screened Coulomb interaction, and explicitly depend
on the sheet-resistance $R_{\square}$ of the 2DEG. These dynamical
effects play an important role in the quantum phase diagram of the
1DJJA. This situation is different from previous approaches in higher
dimensions.\citep{Wagenblast97_SIT_in_a_Tunable_Dissipative_Environment,Vishwanath04_Screening_and_dissipation_at_the_SIT_induced_by_a_metallic_ground_plane}
Indeed, in 1D the plasmon mode is almost statically screened\citep{Lobos10_Dissipative_phase_fluctuations},
and this would lead to the naive conclusion that dynamical effects
are not important. However, a more careful analysis shows that backscattering
originated in the dynamically screened Coulomb potential has the effect
of restoring the SIT. In our system, these dynamical effects have
important consequences for the critical properties of the array, and
should be possible to observe them in dc-transport measurements. Physically,
the coupling to diffusive modes in the metal induces charging effects
which are local in space (i.e., of the order of the lattice parameter
$a$ of the 1DJJA) but which are non-local in time (i.e., Ohmic dissipation
effects), and tend to quench charge fluctuations, rendering superconductivity
weaker. 

By the means of a weak-coupling RG-analysis and a variational approach,
we predict a SIT driven by the presence of dissipation in the 2DEG.
This SIT is of the BKT-type and mediated by unbinding of QPS/anti
QPS pairs, like in the dissipationless case.\citep{bradley_josephson_chain}
Near the critical line the effects of QPS are stronger than those
originated in dissipation and results in a SIT. This scenario is corroborated
by a subsequent variational analysis of action Eq. (\ref{eq:S_eff}),
which suggests the formation of a gap $\Delta$ in the spectrum of
excitations of the 1DJJA {[}cf. Eq. (\ref{eq:g0_combined}){]}. 

Our results suggest that dissipation renormalizes the QPS-rate to
\textit{higher} values and the ratio $\sqrt{E_{J}/E_{0}}$ to \textit{lower}
values {[}cf. Eqs. (\ref{eq:RG_eq_K})-(\ref{eq:RG_eq_eta}){]}, rendering
superconductivity in the 1DJJA weaker. Eventually, an increase of
$R_{\square}$ {[}and therefore of $\eta$, in view of Eq. (\ref{eq:eta_parameter}){]},
could drive the system into the insulating phase, as can be seen in
Figs. \ref{fig:phase_diagram} and \ref{fig:resistivity_eta}. This
phenomenon is different to the case studied by Cazalilla \textit{et
al.}, where QPS processes were absent, and it was dissipation \textit{itself}
that drove the quantum phase transition for the critical value $K_{c}=1/2$.\citep{cazalilla06_dissipative_transition} 

We have also studied the consequences on the temperature-dependent
dc-resistivity of the array $\varrho\left(T\right)$. We have shown
that a non-vanishing $R_{\square}$ induces a rich behavior of $\varrho\left(T\right)$.
In particular in the superconducting phase, where the 1DJJA is in
the Luttinger liquid universality class, and the effects of QPS and
dissipation are perturbative, the resistivity of the array $\varrho\left(T\right)$
follows a power-law behavior $\varrho\left(T\right)=A_{1}T^{\nu_{1}}+A_{2}T^{\nu_{2}}$,
with exponents $\nu_{1}=2K-3$ and $\nu_{2}=2K$ {[}cf. Eq. (\ref{eq:resistivity}){]}
generated by QPS and dissipation, respectively. Therefore, the results
of this paper could be relevant in the interpretation of experimental
results of transport through superconducting circuits subject to dissipative
effects. In the insulating phase, the low-temperature dc-resistivity
is expected to show thermally-activated behavior.\citep{giamarchi_book_1d,fazio01_review_superconducting_networks}
In particular for $K=1$, the resulting model can be studied analytically
with a refermionization approach, and results in a resistivity $\varrho\left(T\right)\sim\eta Te^{\Delta/T}/\Delta$.
Quite importantly, note in this expression that the resistivity depends
also implicitly on $\eta$ via a renormalization of the gap $\eta$.
\begin{acknowledgments}
We acknowledge useful discussions with I. Pop, W. Guichard and F.
Hekking. This work was supported by the Swiss National Foundation
under MaNEP and division II. 
\end{acknowledgments}
\bibliographystyle{apsrev}

\end{document}